# Analysis and Evaluation for the Performance of the Communication Infrastructure for Real Wide Area Monitoring Systems (WAMS) Based on 3G Technology

**M. M. Eissa[1], *Senior Member, IEEE* and Mahmoud M. Elmesalawy[2], *Member, IEEE***

1 Department of Electrical Machine and Power Engineering- Faculty of Engineering - Helwan University – Egypt
2 Department of Electronics, Communications and Computers, Faculty of Engineering - Helwan University – Egypt

***Abstract—*** Wide Area Monitoring Systems (WAMS) utilizing synchrophasor measurements is considered one of the essential parts in smart grids that enable system operators to monitor, operate, and control power systems in wide geographical area. On the other hand, high-speed, reliable and scalable data communication infrastructure is crucial in both construction and operation of WAMS. Universal mobile Telecommunication System (UMTS), the 3G standard for mobile communication networks, was developed to provide high speed data transmission with reliable service performance for mobile users. Therefore, UMTS is considered a promising solution for providing a communication infrastructure for WAMS. 3G based EWAMS (Egyptian wide area Monitoring System) is designed and implemented in Egypt through deployment a number of frequency disturbance recorders (FDRs) devices on a live 220kV/500kV Egyptian grid in cooperation with the Egyptian Electricity Transmission Company (EETC). The developed EWAMS can gather information from 11 FDRs devices which are geographically dispersed throughout the boundary of the Egyptian power grid and to a remote data management center located at Helwan University. The communication performance for the developed EWAMS in terms of communication time delay, throughput, and percentage of wasted bandwidth are studied in this paper. The results showed that the system can achieve successfully the communication requirements needed by various wide area monitoring applications.

## I. INTRODUCTION

Wide Area Monitoring Systems (WAMS) one of smart grid enabling technologies applied on power grid transmission domain to guarantee reliable and efficient power transmission. WAMS involves the use of wide area synchronized measurements; reliable-high bandwidth communication infrastructure; and advanced control schemes [1]. The WAMS combines the data provided by synchrophasor and conventional measurements with the capability of recent communication technologies in order to obtain dynamic information of the entire system [2]. As certain power system measurements cannot be meaningfully combined unless they are captured at the same time, so synchronized measurements technology is considered the core of WAMS. The GPS satellite system is used by the phasor measurement devices to acheive synchronization and timing accuracy for the measurements. Many advanced applications can use advantage of the synchronized measurement capability provided by WAMS as wide area monitoring, real-time operations, improved accuracy of models and forensic analysis, in which wide-area frequency measurements can be used to provide authentication of different media recordings [1–7].

With the fast progress made in the field of synchronized measurement technology and the availability of reliable high-speed communication infrastructures, WAMS becomes possible and practically implemented. For example, a Real Time Dynamic Monitoring System (RTDMS) has been implemented in the Eastern North American bulk power system [8]. Also, a frequency Monitoring Network (FNET) using Frequency Disturbance Recorders (FDRs) is implemented in the North America [9].

The communication infrastructure for WAMS should be carefully developed to provide continuous connectivity between all phasor measurements units in the grid and the control center. Also a huge amount of data from the measurement devices are generated continuously therefore requirements of bandwidth and latency should be met by communication system. Several wired and wireless communication technologies are identified for smart grids. Recent wireless systems offer the benefits of inexpensive products, rapid deployment, low cost installations, wide area coverage, high speed data, and mobile communications compared to wired technologies [4], [5].

Universal Mobile Telecommunications System (UMTS), the 3rd generation cellular network is designed to provide high-speed wireless Internet access and fulfill high quality of service requirements for rapidly growing Internet applications and services. UMTS can support maximum data transfer rates of up to 7.2 Mbit/s in downlink and 384Kbps in uplink [12]. Therefore, UMTS can be considered as a good choice for providing wide area connectivity in WAMS.

The UMTS network consists of three parts: Core Network (CN), UMTS Terrestrial Radio Access Network (UTRAN), and User Equipment (UE). The CN handles routing in the system and stores user information. Since UMTS can be connected to both circuit switched (CS) and packet switched (PS) external network, the CN needs to support both of these switching techniques. The connection between the external networks and CN goes through a Gateway Mobile switching Center (GMSC) in the CS case, and a Gateway GPRS

Supporting Node (GGSN) in the PS case. Furthermore, CS services are routed through a network of Mobile Services Switching Centers (MSCs), and the PS services are routed through a network of Serving GPRS Support Nodes (SGSNs).

The UTRAN consists of a Radio Network Sub-system (RNS), which in turn consists of a Radio Network Controller (RNC) and one or more Node B. Each RNC is connected to both an MSC and an SGSN. The main purpose of the RNC is to manage the radio resources using the Radio Resource Control (RRC) protocol. The Node B is usually called a base station and is the link between user equipment (UE) and the UMTS network. Figure 1, shows the architecture of UMTS system.

.

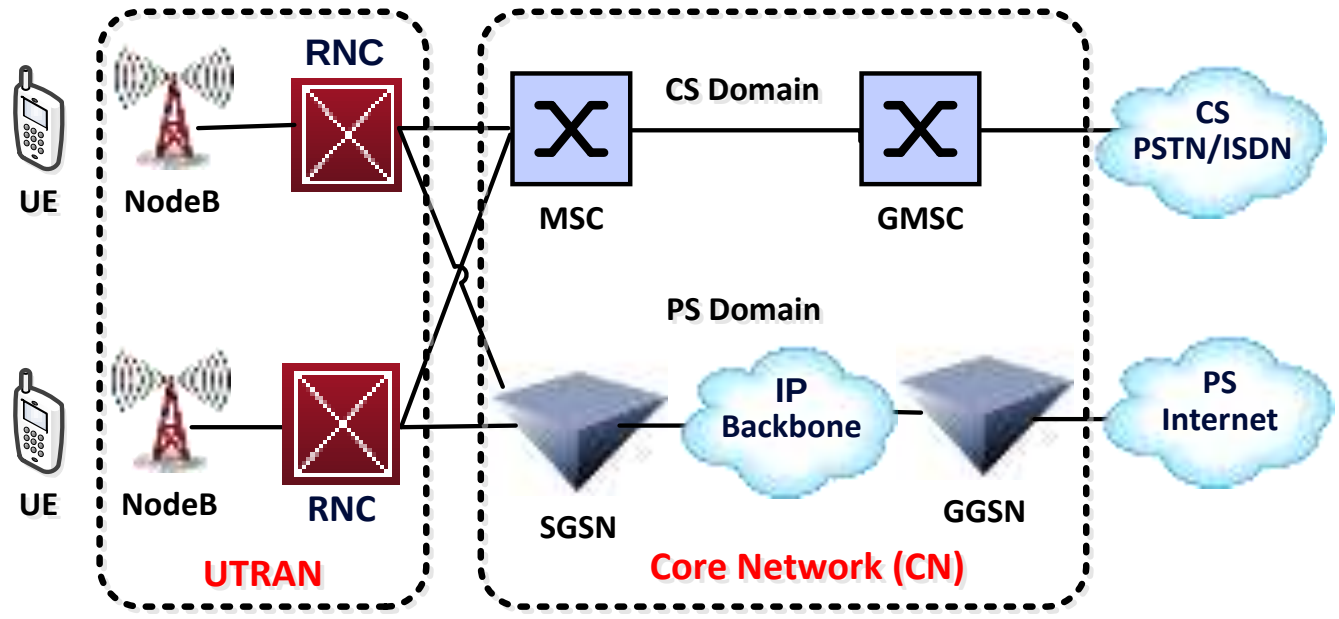

Fig. 1. The Architecture of UMTS.

Using UMTS as communication infrastructure, authors have achieved WAMS based on FDRs devices deployed on Egyptian Power grid for mapping and visualization of system parameters [13], [14]. Ten FDRs devices are deployed on live 220kV/500kV Egyptian grid system in cooperation with the Egyptian Electricity Transmission Company (EETC). The system is implemented as a research project funded from the National Telecommunication Regulatory Authority (NTRA) in Egypt.

In this paper, the performance of the developed EWAMS communication infrastructure is analyzed and evaluated in terms of communication delay, throughput, packet loss, and consumed bandwidth. To the best of our knowledge, our work is the first study of its kind, which evaluates the performance of a live 3G network when it used as a communication infrastructure for WAMS.

The rest of this paper is organized as follows. Section II, introduces the developed EWAMS architecture. Then the implementation and configuration of EWAMS is described in section III. In section IV, the performance analysis methodology and required tools are presented. The experimental architecture and results is provided in section V. Section VI concludes the paper.

## II. EWAMS Architecture

EWAMS is a 3G based smart grid system developed to collect real-time synchronized frequency, voltage, and phase angle measurements at the transmission and distribution levels of the power grid. The EWAMS architecture can be represented by four main building blocks. The first block contains the GPS enabled FDR devices that provide frequency, voltage magnitude, and voltage angle measurements. The second one represents the communication infrastructure that provides the integrated wide area communication media for data measurements transmission. The third block is the remote data management and processing center that provides data gathering, storage, post-disturbance analysis and other information management functions. The last component is the secure remote access connection for different EWAMS online and off-line applications from different remote sectors and clients. Figure 2, shows the EWAMS architecture deployed on the Egyptian grid.

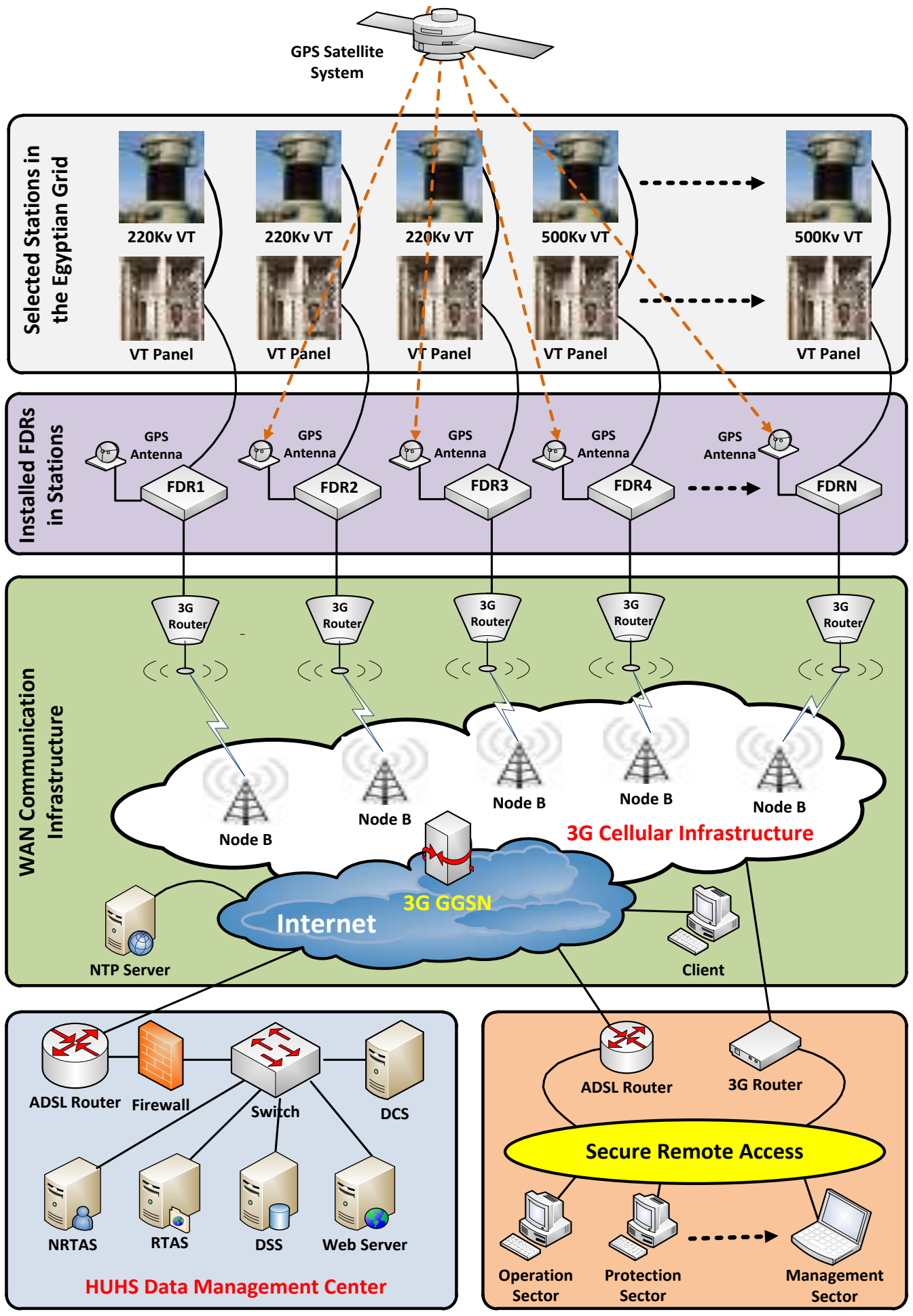

Fig.2. EWAMS Architecture.

In order to discuss the unique characteristics of the EWAMS and evaluate its performance, the main building blocks of the EWAMS are briefly discussed in the following subsections.

### A. *EWAMS Sensors (FDRs)*

As EWAMS is mainly based on the FNET system, FDR device is considered as the key component of the EWAMS. It works as a sensor which performs real time synchronized measurements for frequency, voltage magnitude, and voltage angle with a rate of 10 samples per second and transmits these measurements to a remote data center, hosted in Helwan University for processing and long term storage. Each FDR is equipped with a GPS receiver, which is used to provide the accurate time signal needed for synchrophasor calculation. Also each FDR is embedded with 10/100 Mbps Ethernet interface for IP communication capability.

### *B. HUHS Data Management Center*

Helwan University Host Servers (HUHS) is a data management and utilization system operated by several dedicated servers. The logic behind decomposing the HUHS to several numbers of servers is to distribute the computation power which had the advantage of increasing the system redundancy, reliability, and scalability. The HUHS consists of a number servers connected together through one Gigabit Ethernet local area Network (LAN). The most powerful component of the HUHS system is the data concentrator server (DCS). The primary purpose of the DCS is to collect the measurements transmitted from the FDRs devices and time synchronizes them before distributing these data to different applications for processing and analyzing it.

### *C. EWAMS Communication Infrastructure*

In EWAMS, FDRs devices are distributed over a wide geographic area, covering various locations within the boundary of the Egyptian power system. 3G-UMTS mobile communication infrastructure is used to provide the communication channels between FDRs and DCS in the HUHS data center. 3G router equipped by a 3G-High speed packet access (HSPA) modem is used to allow the FDR to access the Internet through UMTS network as shown in Fig. 3.

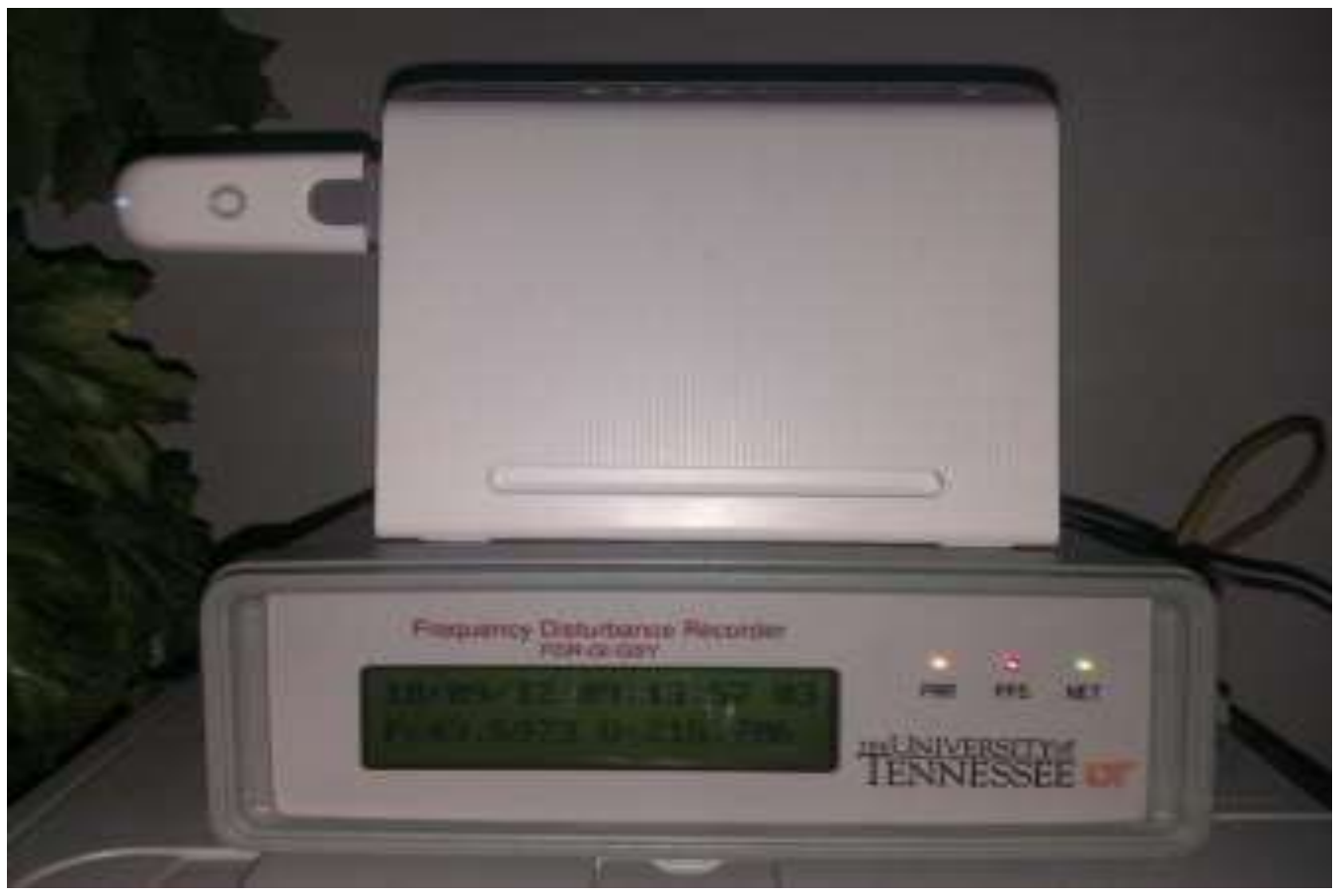

Fig.3. 3G-HSPA modem for FDR Internet connection.

The latest 3G USB modems have an integrated UMTS Subscriber Identification Module (USIM). The USIM contains the subscription information necessary to access the 3G network. They provide wireless access to the Internet using HSPA/HSPA$^+$ technologies with typical maximum download speeds up to 7.2 Mbps and upload speeds 384 kbps.

Client server model is used for the communication between FDRs devices and DCS server. Each FDR device is act as a client and requesting to make a connection with the DCS server. The measurements data is transmitted from FDRs devices to the DCS server using a standard network protocols, Transmission Control Protocol (TCP)/Internet Protocol (IP). IP is fundamental to Internet addressing and routing while TCP is the embodiment of reliable end-to-end transmission functionality which handles packet loss, error control, re-transmission of the lost packets etc.

The TCP connection between each FDR and the DCS is established through a three-way handshake process, ensuring that both FDR and DCS have an unambiguous understanding of the sequence number space. Each FDR sends the DCS an initial sequence number to a predefined destination port, using a SYN packet. The DCS responds with an ACK of the initial sequence number of the FDR in a response SYN packet. Finally, the FDR responds with an ACK of this DCS sequence number and the connection is opened. Figure 4, illustrates the TCP Connection establishment between FDR and DCS.

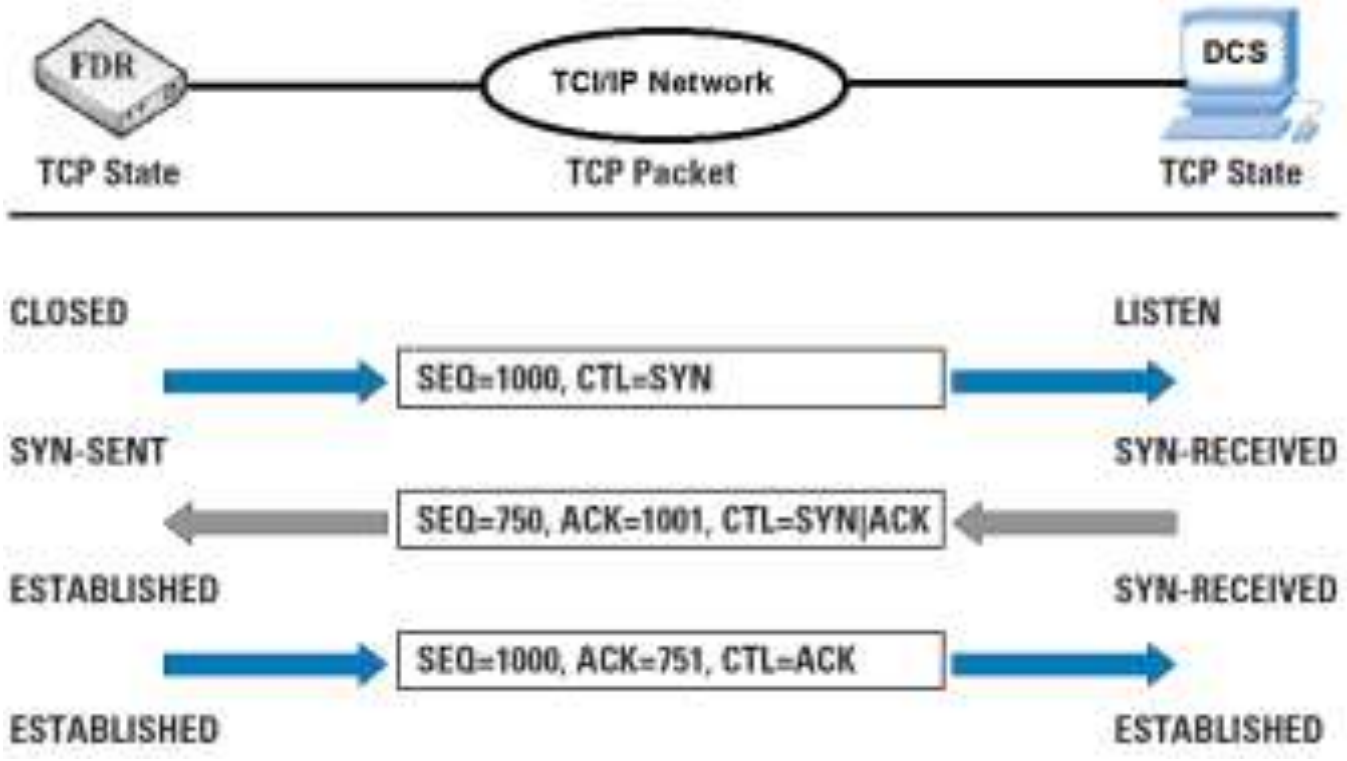

Fig. 4: FDR-DCS TCP Connection establishment.

## III. EWAMS IMPLEMENTATION AND CONFIGURATION

The implemented WAMS system for the Egyptian power grid consists of a number of FDR units geographically dispersed throughout the boundary of the Egyptian power grid, and a data management system (HUHS) located at, Helwan University. The placements of the FDR units are optimally selected to effectively reflect the different frequency coherent areas and to cover as broad an area as possible, in order to capture dynamic behavior of larger system disturbance. Up to now there have been 10 FDRs devices are installed in the power stations over 220kV/500kV transmission level of the Egyptian grid and one FDR is installed on the distribution level at Helwan University. Figure 5, shows the actual locations (latitude and longitude) of the FDRs units that are deployed in the EWAMS system. Due to the minimal installation, FDR units can be easily relocated, if necessary.

Each FDR is configured with the required TCP/IP parameters that enable FDR to send their measurements data to the DCS. These parameters include the destination IP address or domain name for the DCS and the destination TCP port number. The FDR is configured to operate in a TCP client mode which enable the FDR to initiates the connection with the DCS. Figure 6, shows a successfully operated FDR and router in Kurymat power station.

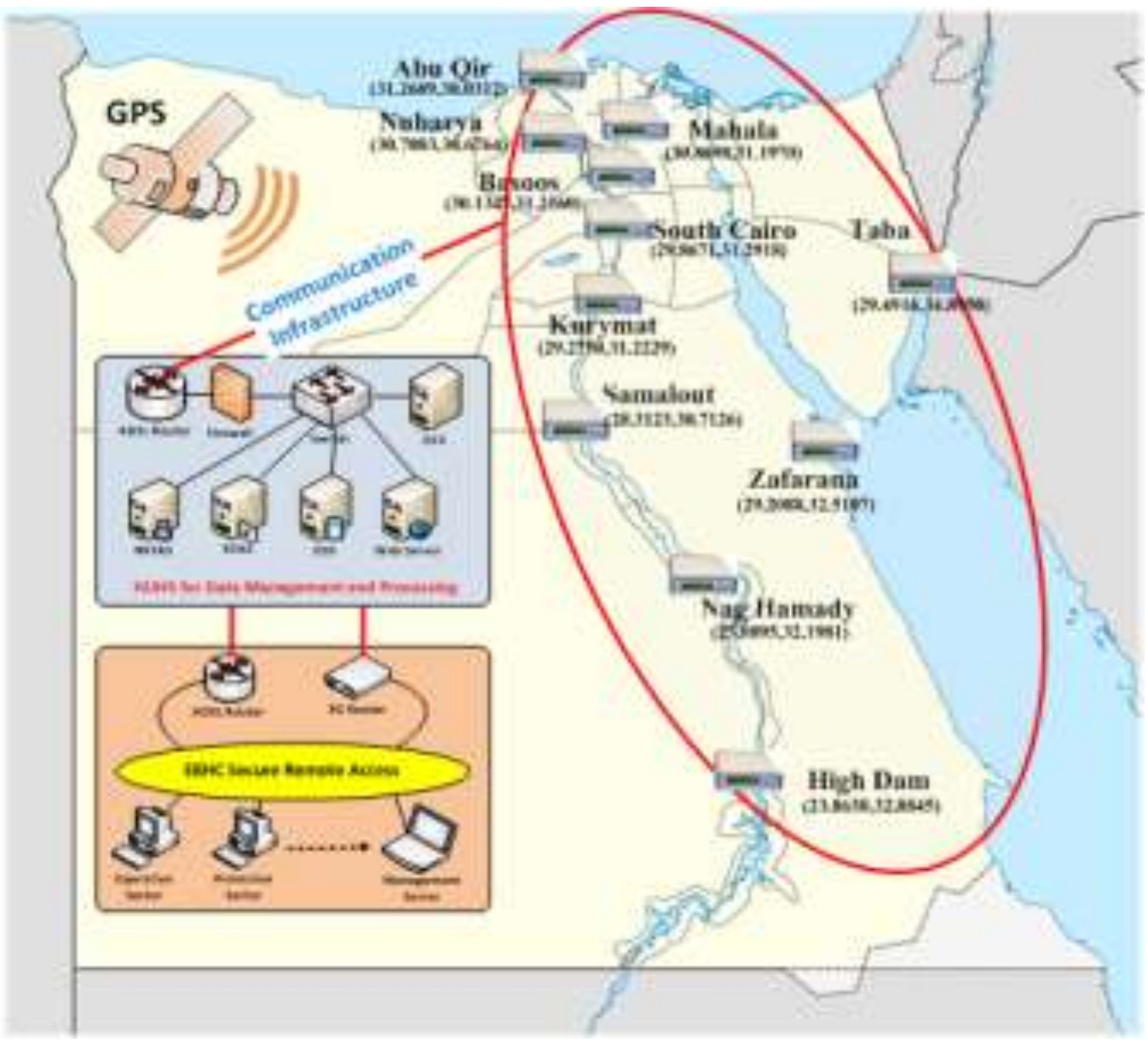

Fig.5. FDRs distribution map on Egyptian 500kV/220kV grid.

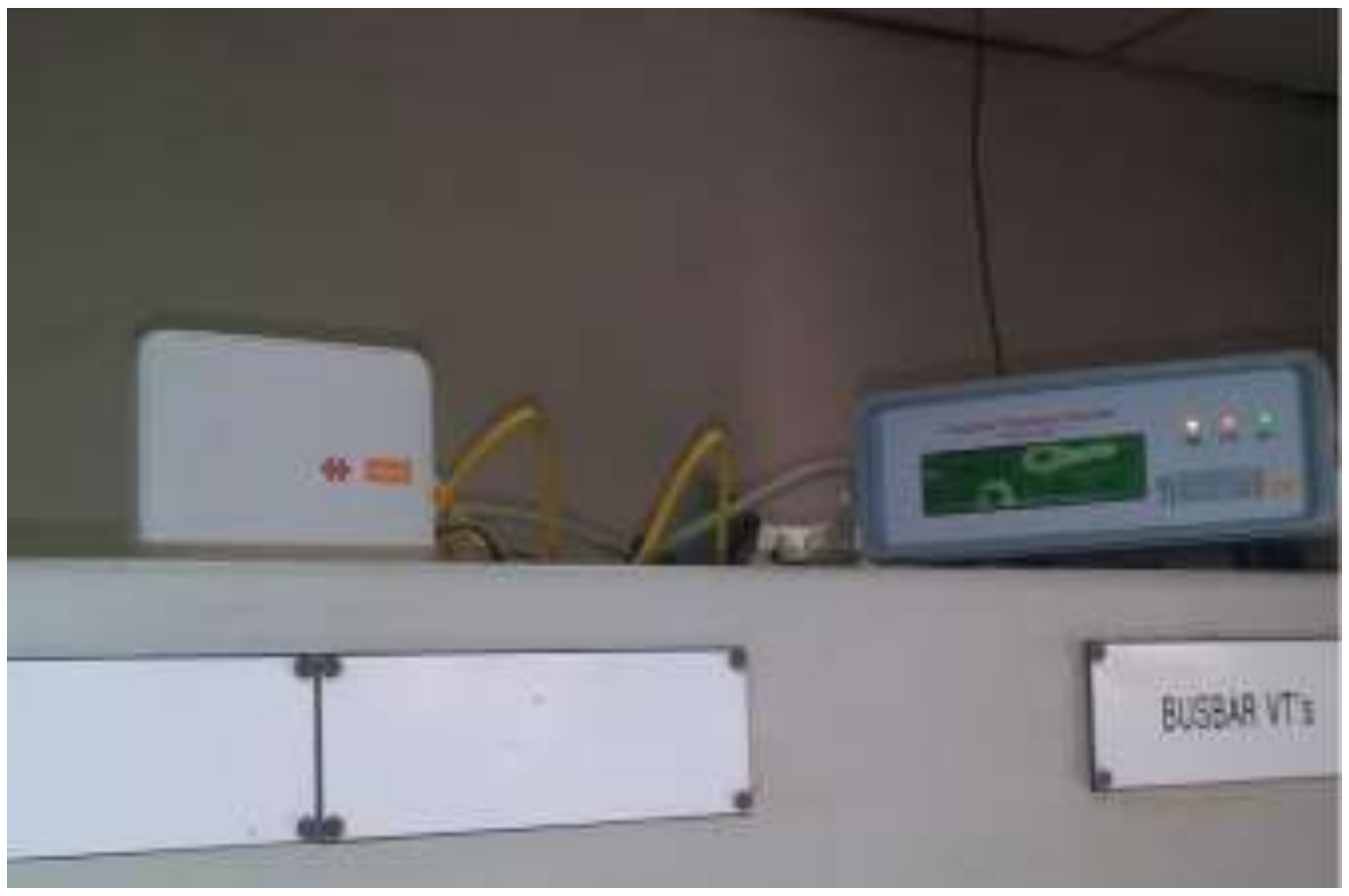

Fig.6. Installed FDR and 3G router in Kurymat power station.

HUHS data center contains physically of four rack mounted servers located at Helwan University and configured to accommodate the different data management and processing functionality required for EWAMS applications. The HUHS center receives data from different FDRs units, processes the data, manages the database, performs data analysis, and supports the web service for the Internet users.

The specifications for each of the four servers are identical with two six core-CPU (Intel Xeon CPU E5-2420 – 2.2GHz), 16GB memory and 2.54 MB cache memory per core and one TB hard disk is employed to meet the need of reliable data transmission, processing and web service. The backup data is stored both on the hard disk of the server and in an external 3TB hard drive. 64-bit Windows Server 2008R2 is the host operating system. Figure 7, shows the four servers and two UPS installed in HUHS data center. The UPS are installed to ensure reliable continuity of servers operation even in case of power blackout.

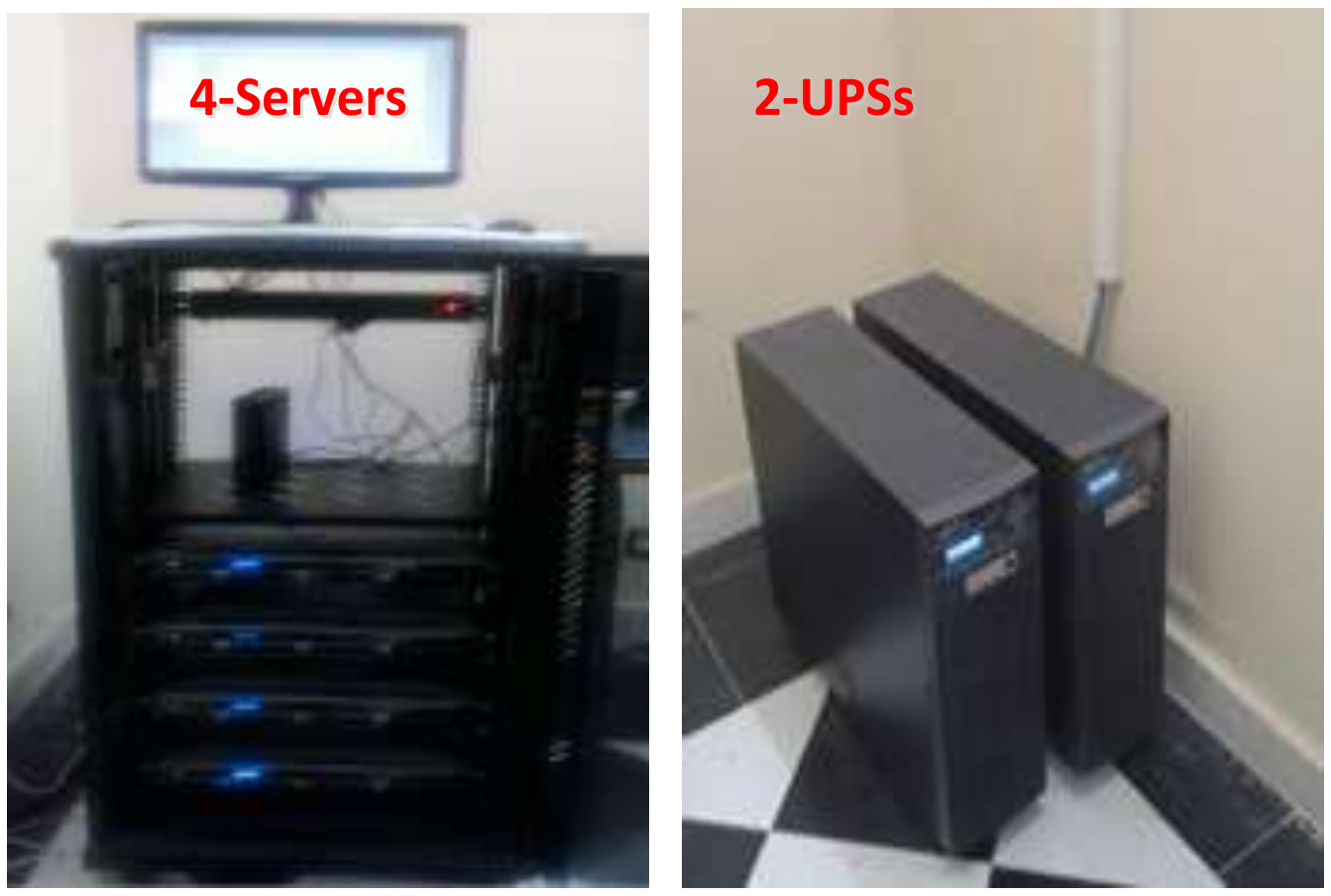

Fig.7. HUHS data center servers and UPSs.

## IV. PERFORMANCE ANALYSIS METHODOLOGY AND TOOLS

In order to evaluate the performance of EWAMS communication infrastructure, analysis methodology and required tools should be firstly specified. Different issues need to be considered: what are the evaluation metrics needed for analyzing the performance of EWAMS communication infrastructure, how much data is needed to have the desired confidence level in the results, and what tools should be used for the performance analysis. These issues will be discussed in the following subsections.

### A. *Performance Evaluation Metrics*

Time delay and throughput are considered the most two critical metrics that should be incorporated into any design or analysis of communication infrastructure for WAMS. The average time required to send one frame from FDR device to the DCS in EWAMS is considered as the average end-to-end (ETE) delay and is expressed as:

$$T_{ETE} = T_{FDR} + T_{CI} + T_{DCS} \qquad (1)$$

Where $T_{FDR}$ represents the fixed delay associated with FDR for phasor measurements calculations, $T_{CI}$ is the average time delay due to communication infrastructure which is defined as the period of time from when a packet is sent out from FDR device until it is received by the DCS, and $T_{DCS}$ is the average processing delay taken by the DCS for decoding the received FDR frames. Since we are interested for nlyzing the performance of communication infrastructure, the time delay due to 3G communication infrastructure can be estimated as follows:

$$T_{CI} = T_p + \frac{L_{FDR}}{R_{UL}} + T_j \qquad (2)$$

Where $T_p$ is the average propagation delay over UMTS network, $L_{FDR}$ is the length of FDR frame which is 55-byte [15], $R_{UL}$ is the uplink transmission rate (bps) in 3G network, and $T_j$ is the associated random delay jitter which accommodates the processing and queuing delay in 3G-NodeB and intermediate nodes as well as the delay due to retransmissions through TCP connections. These delays reflect the viability of the used communication infrastructure, since large delays in monitoring the power system parameters could ruin any control procedures adopted to stabilize the power grid [16]. However, in our analysis we will not break down the time delay into these different parts of the delay as we are interested for the total communication delay.

The second evaluation metric, throughput which is the average data rate of successful frame delivery over a communication infrastructure (This is also referred to as goodput). It is usually measured in bits per second (bps) and sometimes in data packets per second or data packets per time slot. Actually, the communication delay and throughput are two dependent variables, as delay spikes on 3G based Internet communication can cause spurious TCP timeouts leading to significant throughput degradation. The problem is that delay on Internet connections is highly variable resulting for instance from route flipping [17]. On the one hand, underestimation of round trip delay (RTT) leads to a premature retransmission timeout in case there is no loss or the retransmission could be handled by the fast retransmission mechanism. On the other hand, overestimation of RTT leads to a delayed retransmission timeout, in case there is a loss that cannot be captured by the fast retransmission mechanism. Therefore, the percentage of TCP retransmissions and fast retransmissions are also studied in this paper in order to evaluate the suitability of the selected value of retransmission time out (RTO) settled on the FDRs devices in EWAMS architecture.

### B. *Sample Data Size for Performance Evaluation*

As each FDR device send their data to the DCS with a rate of 10 frames per second continuously (24 hour per day), so a huge amount of data is received by the DCS from each FDR that make it difficult to be analyzed for a long time interval. So in order to have the desired confidence level in the analysis results, the minimum significant amount of data samples that need to be used for analysis should be determined. The confidence level indicates how sure of the results we can be. It is expressed as a percentage and represents how often the true percentage will fall within a specified confidence interval. The minimum size of sample data required for analysis can be determined in two steps: The first one is pre-sampling to calculate the standard deviation for evaluation metric x as follows.

$$S_x = \sqrt{\frac{\sum_{n=1}^{N_P}(x_n - \bar{x})^2}{N_P}} \qquad (3)$$

Where $N_P$ is the pre-sample size, $x_n$ is the $n^{th}$ sample, and $\bar{x}$ is the mean of the selected samples. In the second step, the calculated standard deviation and the required confidence level are used to calculate the minimum number of samples required to measure the metric $x$ as in (4) [18].

$$N_{min}^{x} = \frac{Z^2 . {S_x}^2}{e^2 + \frac{Z^2 . {S_x}^2}{N_T}} \qquad (4)$$

Where $N_T$ is the total population size, $e$ is the acceptable standard error that can be set to meet our analysis demand, and $Z$ is called Z-score which is determined according to the required confidence level (ex. $Z$ has a value of $1.959$ for a confidence level 95%) [19].

Let $S_D, and\, S_{Th}$ represents the standard deviations calculated using the pre-samples set for the two evaluation metrics, communication delay and throughput respectively. Then, given, the required confidence level and the acceptable standard error $e$, the minimum significant number of samples required to be collected to measure each metric can be determined using (4). If $N_{min}^{D}$ and $N_{min}^{Th}$ represents the minimum number of samples required to measure the communication delay and throughput respectively, then the minimum number of samples required to be collected in our experiment can be determined as follows:

$$N_{min}^{D,Th} \geq Max.\left(N_{min}^{D}\, , N_{min}^{Th}\right) \qquad (5)$$

A simple random sample of size $N_{min}^{D,Th}$ from the total population size $N_T$ chosen in such a way that every set of $N_{min}^{D,Th}$ samples has an equal chance to be the sample actually selected, which is ideal for statistical purposes.

### C. *Tools for Analysis*

For the experimental design, we must select appropriate tools and applications. A good traffic capturer and analyzer play a core role throughout data collection and analysis processes. How to ensure time synchronization should be considered when calculating the one way communication time delay from each FDR device to the DCS.

Wireshark [20], one of the world's foremost network protocol analyzer, and is the defacto standard across many industries and educational institutions is used as a packet analysis tool in our experimental work. Wireshark is configured to capture the packets received on the DCS network interface that are sent by FDRs devices, decodes them and presents them in an easy to understand format. Protocol Dissector is used by wireshark which allows it to break down a packet into various sections so that it can be decoded and analyzed.

As in this research, we would like to calculate the communication time delay for packet transmission from FDR devices to DCS, we have to guarantees that both FDRs and DCS are synchronized to same clock, thus time synchronization need to be configured on both sides. The network time protocol (NTP) is a protocol designed to synchronize the clocks of computers over a network. NTP is used to synchronize the system time on the DCS with the universal time coordinated (UTC) time. On the other hand, each FDR device is developed to send their measurements in the form of time-stamped packets that synchronized to UTC time by using its embedded GPS receiver.

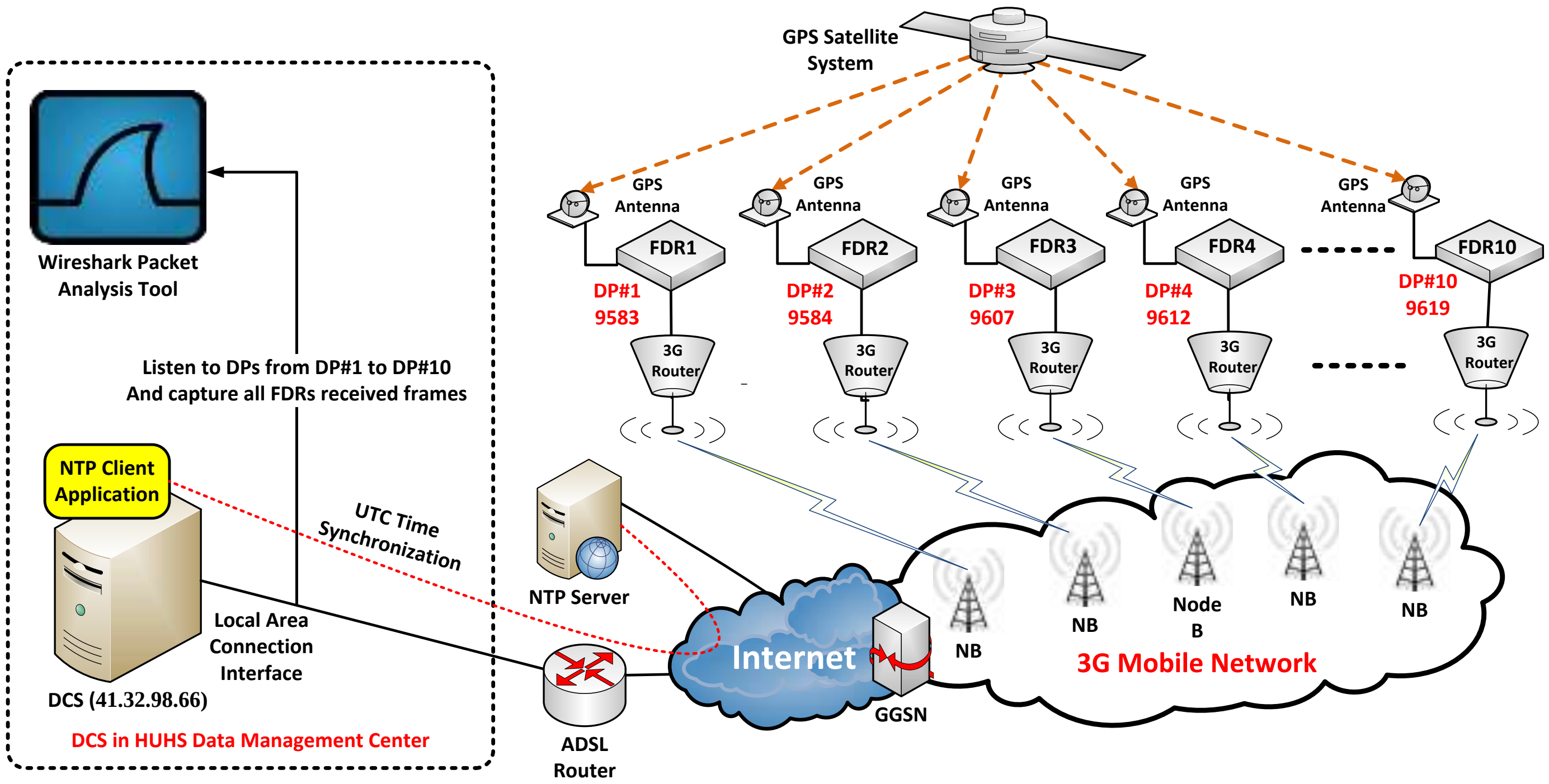

| Power Station | Zafarana | Nubaraya | Nag-Hamady | High-Dam | Mahala | Samalot | Abu-Kir | S. Cairo | Kurymat | Bassos |
|---|---|---|---|---|---|---|---|---|---|---|
| FDR Device No. | FDR#1 | FDR#2 | FDR#3 | FDR#4 | FDR#5 | FDR#6 | FDR#7 | FDR#8 | FDR#9 | FDR#10 |
| Destination Port No. | 9583 | 9584 | 9607 | 9612 | 9613 | 9614 | 9615 | 9616 | 9617 | 9619 |

Fig. 8. Experimental Setup Architecture.

Since both FDRs devices and DCS are time synchronized, the packets arrival time displayed on Wireshark became accurate and useful for determining the communication time delay between FDRs and DCS. When FDR sends a packet 'x' with a time stamp $T_{Send}$ that inserted by the embedded GPS receiver on the FDR. Then the corresponding time when packet 'x' arrives at the destination interface of the DCS can be got as $T_{Receive}$. Therefore the time delay due to communication infrastructure can be directly obtained as $(T_{Receive} - T_{Send})$. Given the time delay of each packet, we can analyze the time delay distribution for each FDR device and characterize this distribution's average value as a result.

## V. EXPERIMENTAL ARCHITECTURE AND RESULTS

In this section, the experimental architecture for analyzing and evaluating the performance of EWAMS communication infrastructure is presented. Then the results are showed and discussed. In EWAMS, the FDRs devices are configured to operate in a TCP client mode which enable the FDR to initiates the connection with the DCS through three way handshaking procedures. Each FDR is configured with the required TCP/IP parameters that enable FDR to send their measurements data to the DCS. These parameters include the destination IP address for the DCS and the destination TCP port number. The source IP address of the FDR is dynamically allocated by the connected 3G router. On the other hand, The DCS server is configured with static IP address (41.32.98.66).

NTP client application is installed on the DCS server which communicates with an NTP server through Internet to synchronize the DCS system time with the UTC time. Many of servers providing NTP service are available on the Internet. "Server0.africa.pool.ntp.org" is selected as the NTP server for our time synchronization. Wireshark is installed on the DCS server and configured to listen and captures all packets received on its local Ethernet interface. Fig.8 shows the experimental architecture for evaluating the EWAMS communication infrastructure.

In order to obtain statistical results represents the performance of the communication infrastructure on different time periods having different traffic loads, one complete day time interval is used in our analysis. Since one day time interval represents 86400 seconds, if we choose one second to be our basic unit of time which includes 10 data samples from each FDR device.

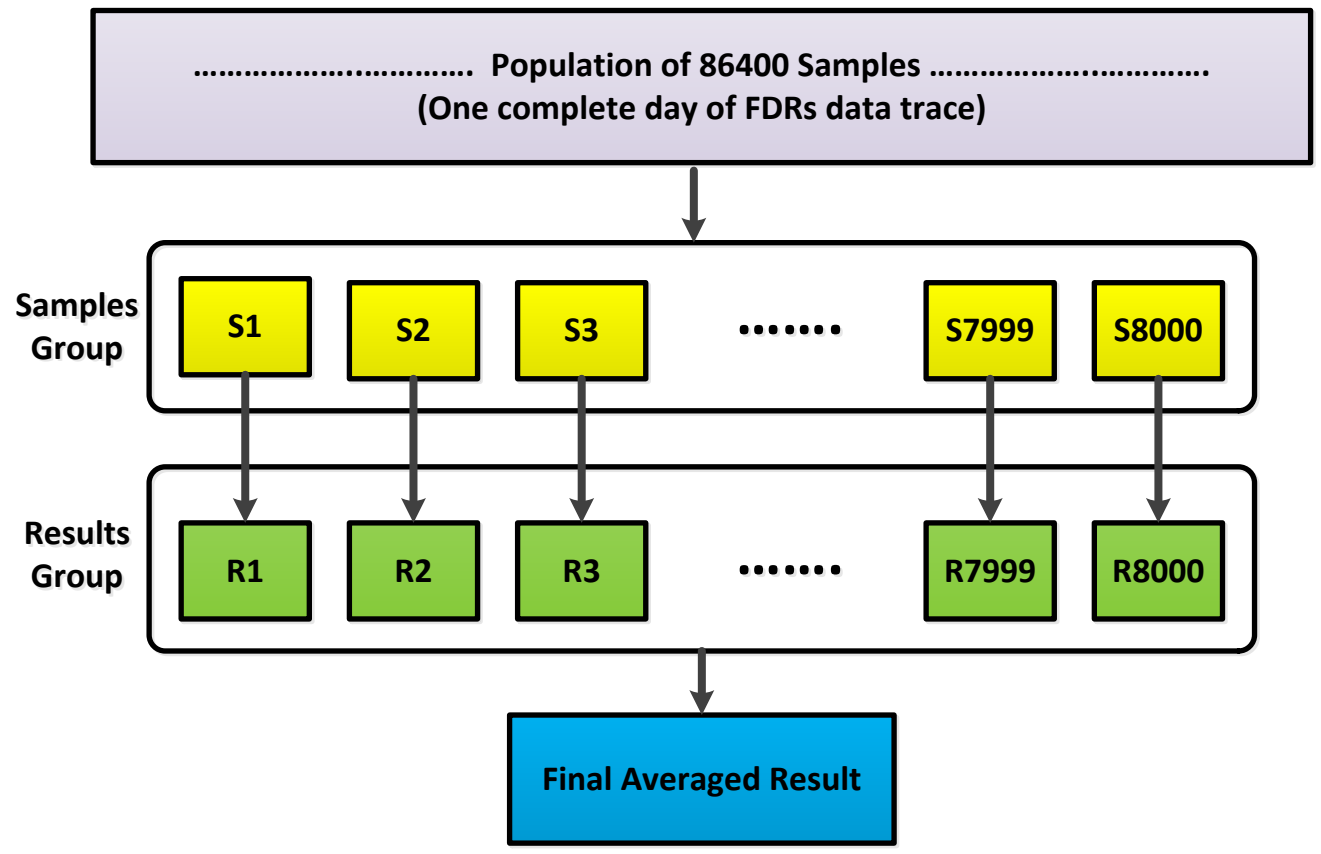

Fig .9. Sampling, data collection and analysis procedure for each FDR device.

Pre-samples of size 10000 are used to estimate the standard deviations for communication delay and throughput metrics averaged over all FDRs devices. The estimated values for the two metrics are 0.908 and 0.186 for $S_D, and\ S_{Th}$ respectively. Given, 95% confidence level which is more widely used by researchers [18], and standard error $e = 2\%$, the minimum number of samples required to be collected to measure the communication delay and throughput are calculated using (4) to get $N_{min}^{D} = 7246.63$ and $N_{min}^{Th} = 330.65$. Therefore, the minimum number of samples required to be collected in our experiment can be chosen with any value greater than or equal $7246.63$. This result indicates that we have to pick at least 7247 samples from the population of 86400 to reach the 95% confidence level. So, data samples of size 8000 are collected and are used in our analysis. Figure 9, shows the whole procedures for each FDR device.

Based on the above results, Wireshark is executed to capture the received frames from 10 installed FDRs devices on the power stations for one complete day time interval. Then 8000 data samples that are corresponding to 8000 seconds (≈2.2 hours) interval are randomly selected with equal chance from the whole 86400 population samples. Figure 10, shows an example of the log file when using the installed Wireshark on DCS to capture the packets sent by FDRs devices.

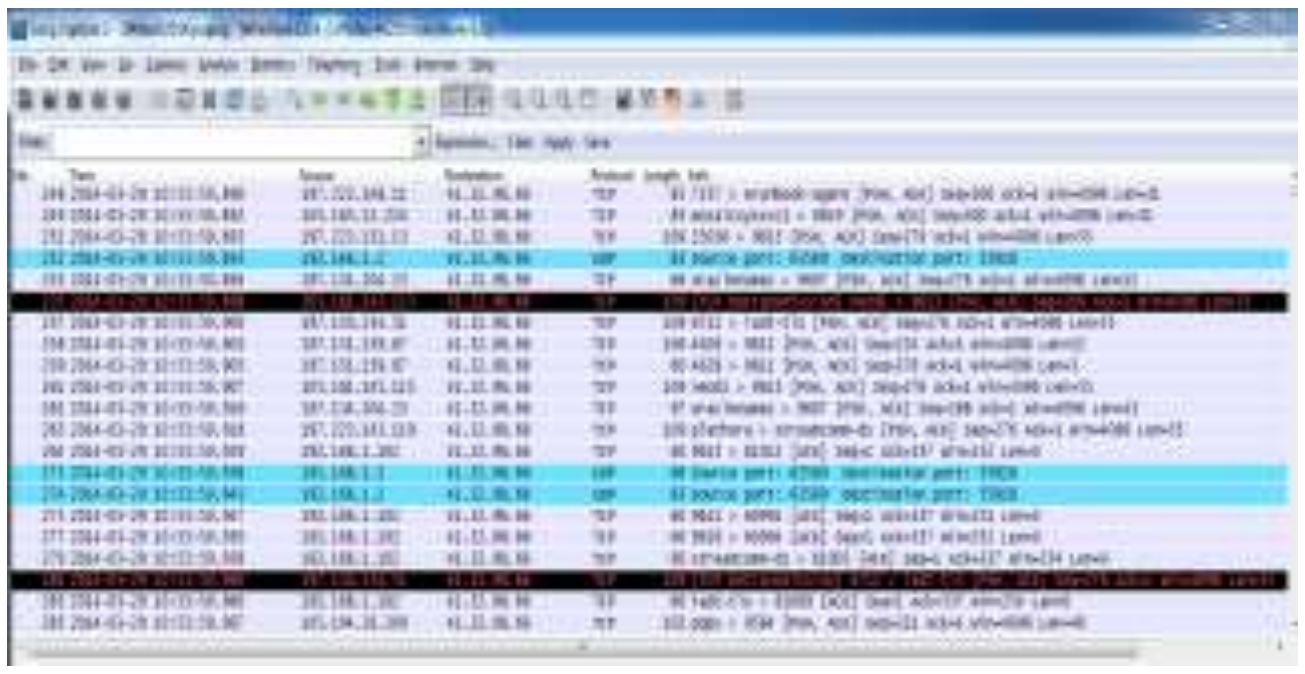

Fig. 10. Wireshark log file for FDRs captured packets.

Figure 11, shows a snapshots from the TCP conversations between DCS and two FDRs devices #3 and #6. As can be noted, in some times, the FDR frame of length 55 bytes is transmitted in more than one TCP segments. This will lead to degradation in the gross throughput and protocol efficiency for FDRs devices in which additional amount of bytes (TCP and IP headers) need to be added for each part of transmitted data from the FDR frame.

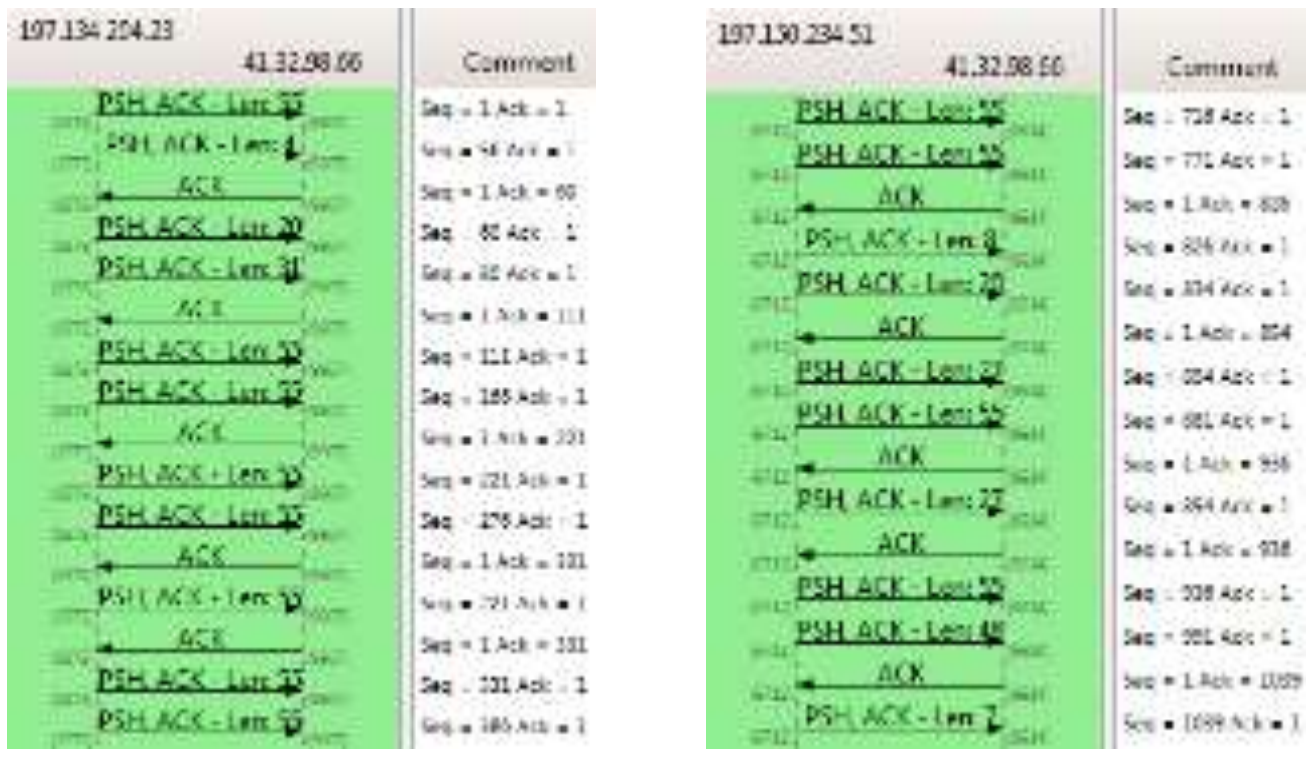

(a) FDR Device#3 (b) FDR Device#6

Fig. 11. FDR-DCS TCP Conversations for two FDRs devices.

An also important consideration is gained from these graphs is that the receive time for each FDR data sample is measured at the time of receiving the last part of the FDR frame (i.e. the time when receiving complete FDR frame with length of 55bye) and not the time of receiving each TCP segments. This is an important issue for getting accurate results for communication delay calculations.

Figure 12, shows the measured throughput for four FDRs devices plotted for the first 200 samples from the selected set of samples. Throughput is estimated per each selected sample by dividing the total amount of bytes successfully transferred by the capture duration and it's calculated in one direction. As can be seen, the throughput varies with a relatively large range (from approximately 8Kbps to 18 Kbps). High peaks indicate that the throughput of the particular flow had high bandwidth utilization. This variation reflects the instantaneous change in channel condition and transmission delay over UMTS-based Internet connections. Poor channel conditions and exceed time delay may lead to large number of packet retransmissions that will degrade the throughput. It is also important to note that there is a variation in the behavior of throughput between different FDRs devices. This can be explained by the different effect produced from the cellular Internet users on FDR data transmissions in each area. Since FDRs devices are geographically distributed over wide area from the south of Egypt to its north, so each FDR device is actually served by different eNodeB with different load of cellular Internet users.

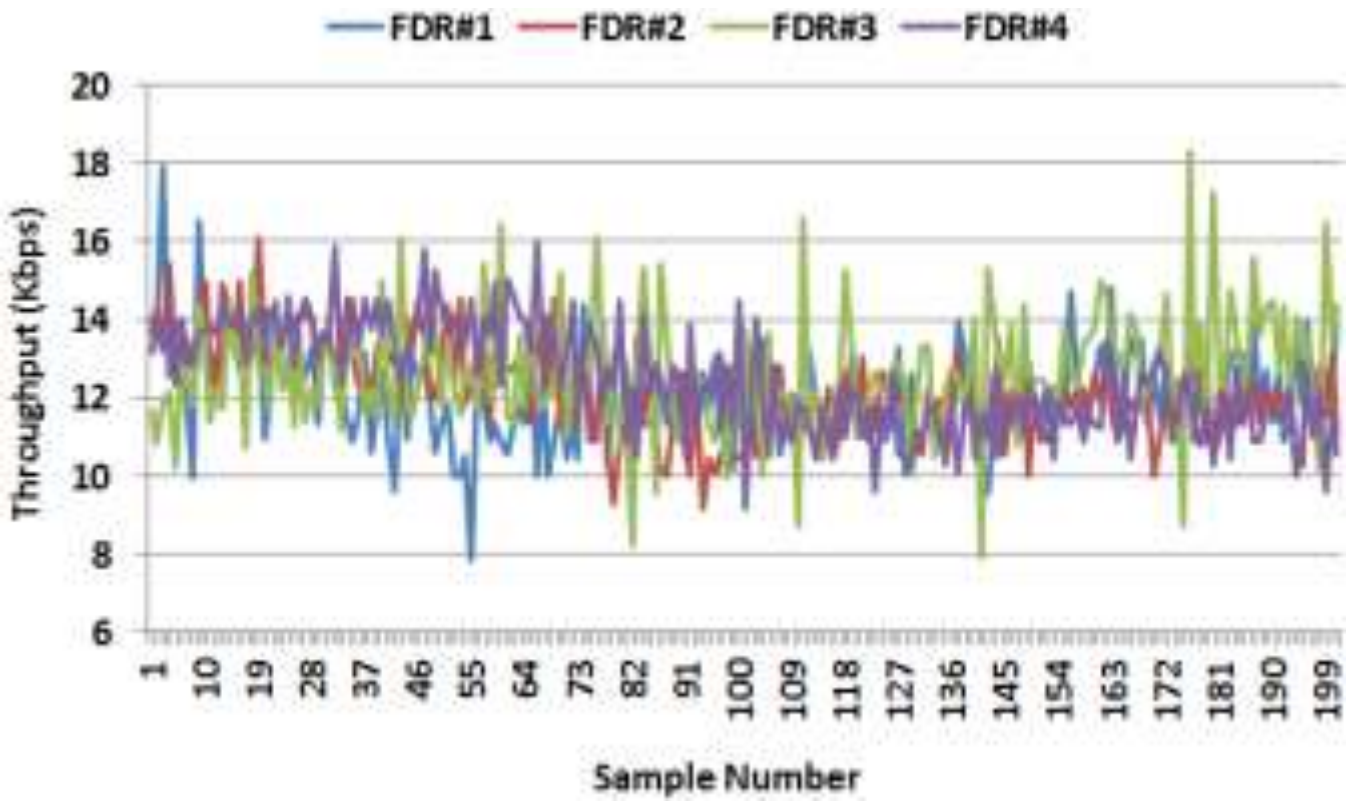

Fig. 12. Measured throughput for different FDRs devices.

The percentages of TCP retransmissions and fast retransmissions for each FDR device are depicted in Fig. 13. These values are calculated as the percentage ratio of the number of retransmitted bytes over the total number of bytes transmitted during the measurement tests. TCP retransmission occurs when the FDR device retransmits a packet after the expiration of the acknowledgement time out timer. On the other hand, TCP fast retransmission occurs when the FDR device retransmits a packet before the expiration of the acknowledgement timer. This is because FDR receive some packets which sequence numbers are bigger than the acknowledged packets. So FDR fast retransmit upon receipt of three duplicate ACKs. As can be seen, all FDRs devices have a TCP retransmission rate less than 0.3% which is considered to be acceptable. The average value over all FDRs is 0.27% with a standard deviation of 0.02. On the other hand, very low percentages of fast retransmissions with maximum value of 0.008% are estimated for different FDRs devices as shown in Fig. 13(b).

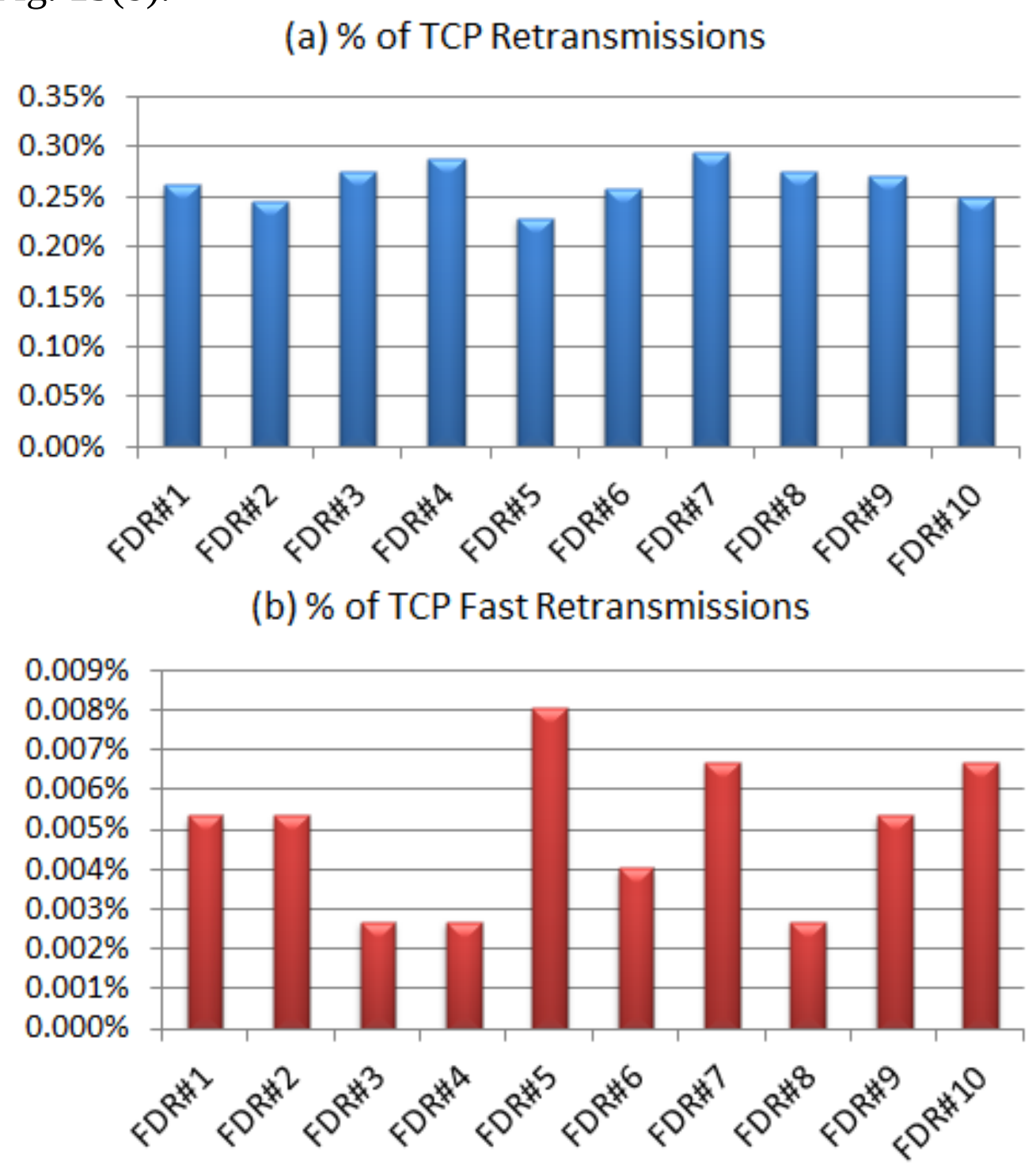

Fig. 13. Percentage of TCP Retransmissions.

.

The number of TCP retransmissions is considered the main source of wasted amount of bandwidth and has a great effect on the throughput. Figure 14, depicts the bandwidth wasted due to retransmissions as a percentage of the flows. As seen, 2.703% of the total available bandwidth is wasted as a result of TCP retransmissions and fast retransmissions.

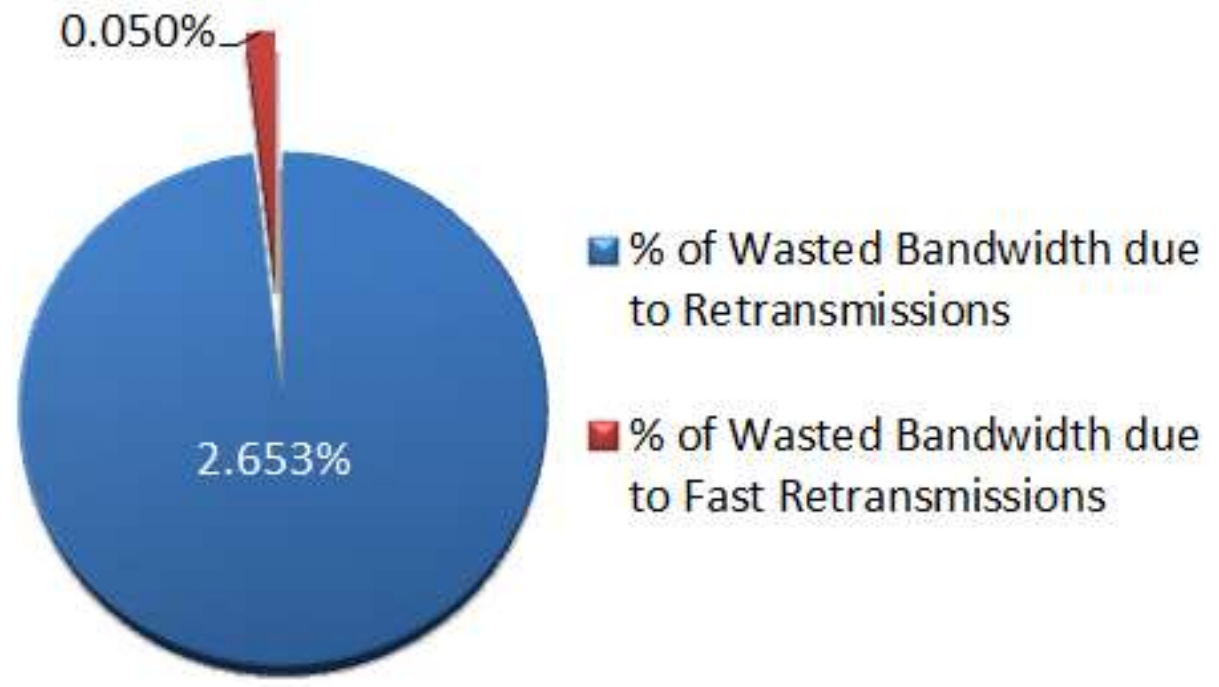

Fig. 14. Bandwidth Wasted Due to Retransmissions.

Figure 15, shows the average one way communication delay for four FDRs devices for first 200 samples from the selected set. As can be seen, the delay due to communication infrastructure ranges approximately from 100ms to 170ms. This instantaneous variation in the average communication delay can be explained by the different load on different NodeBs associated to different FDRs devices in each area as well as the different distances between each FDR device and its associated NodeB that lead to different transmission delays. Also the different load on Internet at different time intervals is a key factor that leads to this variation.

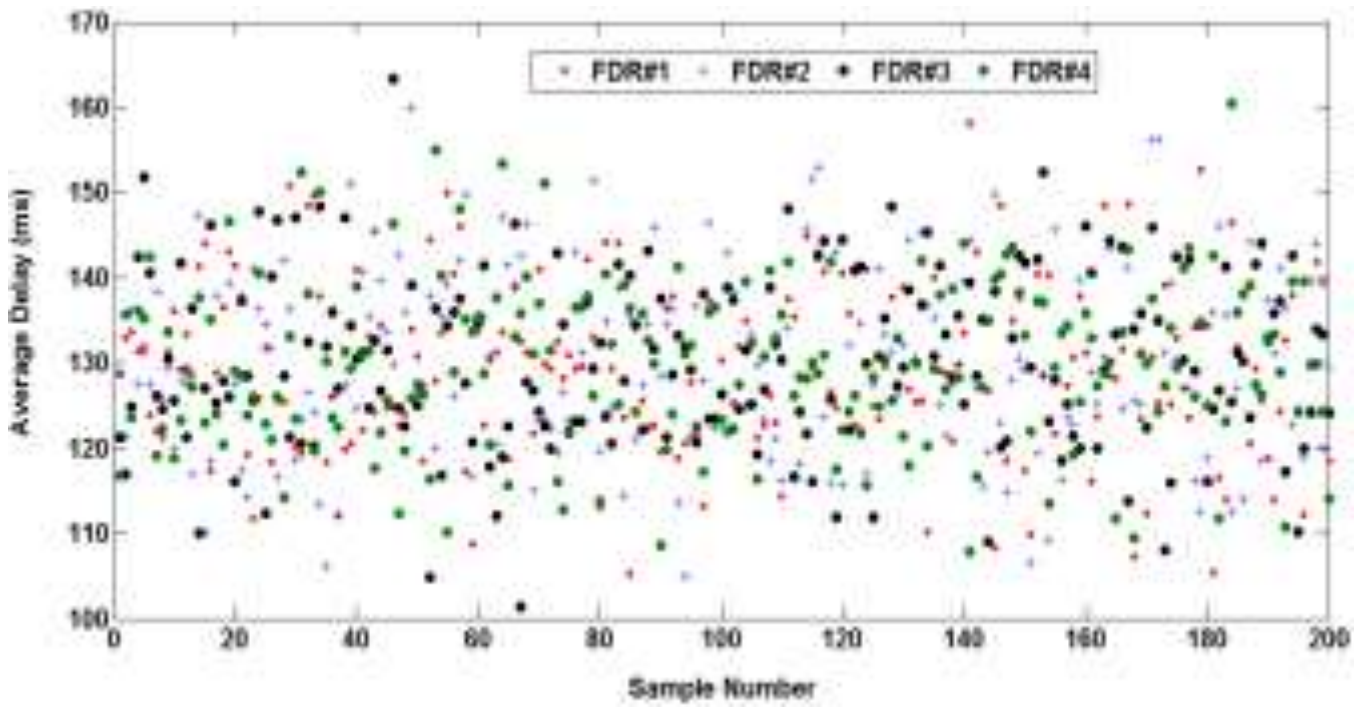

Fig.15. Scatter diagram of one way time delay.

Table I, summarize the values of different evaluation metrics averaged over the number of selected samples. As shown, the average throughput is around 12Kbps for all devices, the average communication delay ranges from 128.986 ms to 131.627ms with a maximum deviation equal 68.361, while the percentage of wasted bandwidth ranges from 0.237% to 0.301. This implies that the communication performance of different devices in terms of the mentioned evaluation metrics is approximately the same over long time intervals.

Table I. Evaluation metrics values for different FDRs devices.

| FDR Device Number | Average Throughput (Kbps) | Average Communication Delay (ms) | % of Wasted Bandwidth |
|---|---|---|---|
| FDR#1 | 12.044 | 129.450 | 0.268 |
| FDR#2 | 12.146 | 129.983 | 0.252 |
| FDR#3 | 12.634 | 130.296 | 0.279 |
| FDR#4 | 12.356 | 129.760 | 0.292 |
| FDR#5 | 12.561 | 131.547 | 0.237 |
| FDR#6 | 12.254 | 128.986 | 0.262 |
| FDR#7 | 12.432 | 129.342 | 0.301 |
| FDR#8 | 12.612 | 129.675 | 0.279 |
| FDR#9 | 12.367 | 130.352 | 0.276 |
| FDR#10 | 12.148 | 131.627 | 0.257 |

As the maximum instantaneous delay value for any of the FDRs devices doesn't exceed 183.216ms. And hence, the acceptable communication delay for WAMS monitoring applications should be $\leq 1000ms$ [21], so UMTS communication infrastructure can be considered a good choice for developing WAMS based monitoring applications.

## VI. CONCLUSION

A real Wide Area Monitoring System on 220Kv/500kV Egyptian grid is developed using 10 FDRs devices in cooperation with the Egyptian Electricity Transmission Company for installing the devices. The system succeeded to monitor the dynamic information about the Egyptian power system in a real time covering many areas on the Egyptian grid. 3G-UMTS mobile communication infrastructure is used to provide the required communication channels between FDRs devices and the data management center located at Helwan University. The communication performance for the developed WAMS is analyzed and evaluated. Different parameters are considered in the analysis process includes; communication time delay, throughput, and percentage of wasted bandwidth are studied in this paper. The results showed that the developed system can achieve successfully the communication requirements needed by various wide area monitoring applications. Hence, as a final conclusion, we can say that UMTS mobile communication networks can be considered as a good choice for developing WAMS based monitoring applications.

## ACKNOWLEDGMENT

The authors gratefully acknowledge funding and support from **National Telecom Regulatory Authority (NTRA)**, Egypt (http://www.ntra.gov.eg) to implement the network architecture proposed in this work. Also, we would like to thank Egyptian Electricity Transmission Company (EETC) for help in implementing the FDRs devices on real system. More details about the project progress can be obtained on (www.helwan-ntra.com).